\documentclass[prb,reprint,superscriptaddress,showkeys,twocolumn,amsmath,amssymb]{revtex4-2}
\usepackage{graphicx}
\usepackage[colorlinks=true,allcolors=blue]{hyperref}
\usepackage{braket}

\newcommand{\Q}{\mathcal{Q}}
\newcommand{\m}{\mathfrak{m}}
\newcommand{\cm}{\mathfrak{c}}

\date{\today}

\DeclareGraphicsExtensions{.pdf,.ai}
\DeclareMathOperator{\Tr}{Tr}

\begin{document}


\title{Counting interacting electrons in one dimension}

\def\afflux{Department of Physics and Materials Science, University of Luxembourg, L-1511 Luxembourg, Luxembourg}
\def\affjul{Peter Gr\"{u}nberg Institute, Theoretical Nanoelectronics, Forschungszentrum J\"{u}lich, D-52425 J\"{u}lich, Germany}

\preprint{APS/123-QED}

\author{O. Kashuba}
\email[Email: ]{o.kashuba@fz-juelich.de}
\affiliation{\affjul}
\author{T. L. Schmidt}
\affiliation{\afflux}
\affiliation{School of Chemical and Physical Sciences, Victoria University of Wellington, P.O. Box 600, Wellington 6140, New Zealand}
\author{F. Hassler}
\affiliation{Institute for Quantum Information, RWTH Aachen University, 52056 Aachen, Germany}
\author{A. Haller}
\affiliation{\afflux}
\author{R.-P. Riwar}
\affiliation{\affjul}


\begin{abstract}
The calculation of the full counting statistics of the charge within a finite interval of an interacting one-dimensional system of electrons is a fundamental, yet as of now unresolved problem.
Even in the non-interacting case, charge counting turns out to be more difficult than anticipated because it necessitates the calculation of a nontrivial determinant and requires regularization.
Moreover, interactions in a one-dimensional system are best described using bosonization.
However, this technique rests on a long-wavelength approximation and is a priori inapplicable for charge counting due to the sharp boundaries of the counting interval.
To mitigate these problems, we investigate the counting statistics using several complementary approaches.
To treat interactions, we develop a
diagrammatic approach in the fermionic basis, which makes it possible to obtain the cumulant generating function up to arbitrary order in the interaction strength.
Importantly, our formalism preserves charge quantization in every perturbative order.
We  derive an exact expression for the noise and analyze its interaction-dependent logarithmic cutoff.
We compare our fermionic formalism with the results obtained by other methods, such as the Wigner crystal approach and numerical calculations using the density-matrix renormalization group.
Surprisingly, we show good qualitative agreement with the Wigner crystal for weak interactions, where the latter is in principle not expected to apply.
\end{abstract}

\maketitle

\section{Introduction}

The full counting statistics (FCS) of an observable collects the information about its quantum measurement in a single function~\cite{Levitov1996}.
It is particularly useful for charge counting in one-dimensional systems, where it provides a compact representation of transport properties~\cite{Schoen2006,Mirlin2010}.
Furthermore, it is closely related to the entanglement entropy, which can be written in terms of the even cumulants in systems that can be mapped onto non-interacting fermions~\cite{LeHur2012}.
Importantly, the FCS can reveal intricate properties of observables which may remain hidden in low cumulants.
For instance, the moment generating function $\m(\lambda)\equiv\langle e^{i\lambda \Q}\rangle$ of the number of particles $\Q$ in a given interval $l$ contains information about charge quantization, which manifests itself in the global symmetry $\m(\lambda+2\pi)=\m(\lambda)$ and $\mathrm{Im}[\m(\pi)]=0$.
Here $i\lambda$ plays role of a purely imaginary counting field.
In contrast, charge quantization is not evident in the average particle number $\braket{\Q}$ or its fluctuations $\braket{\Q^2}$ alone.
The moment generating function also plays an important role for the spin correlations in spin-$1/2$ Heisenberg chains because at $\lambda=\pi$ it has the same form as a Wigner-Jordan string factor contained in, e.g., the form of the spin raising operator $\sigma_{j}^{+} = e^{i\pi \sum_{k=1}^{j-1}c_{k}^{\dag} c_{k}} c_{j}^{\dag}$~\cite{Montroll1963,Luther1975,Shelton1996}.

Charge quantization is a particularly interesting issue in interacting systems.
In low-dimensional electronic systems with strong correlations, effective low-energy field theoretical treatments have demonstrated a remarkable success by invoking emergent excitations carrying a only a fraction of the elementary charge.
First pioneered by Jackiw and Rebbi for a relativistic fermion-soliton model~\cite{Jackiw_1976}, this peculiar notion also appeared subsequently in solid-state systems, be it for the Su-Schrieffer-Heeger model~\cite{SSH_1979}, the fractional quantum Hall effect~\cite{Laughlin_1983,Moore_1991,Kane_1994,Saminadayar_1997,de-Picciotto_1997,Stern_2008}, or in the Tomonaga-Luttinger liquid~\cite{Pham_2000,Mirlin2010}.
For condensed-matter systems with a well-defined vacuum state however, the notion of fractional charges is only meaningful on sufficiently large length scales, implying a certain ``fuzziness'' of the charge observable~\cite{Haldane_1981,Rajaraman_1982,Kivelson1982,Riwar_2021}.
It is therefore an interesting question to understand the interplay between effective fractional charges in correlated systems and the fundamental elementary charge, observed in the FCS, by increasing the spatial resolution of a given charge detector.
However, as it turns out, already for the generic model of interacting electrons in 1D, this is a very hard problem because the standard bosonization technique requires an artificial removal of the lower bound in the spectrum~\cite{Mattis1965}, and is thus simply not capable of answering questions of this type~\cite{Haldane_1981}.
In this work, we provide a first step towards this goal by developing a novel diagrammatic technique to compute the moment and cumulant generating functions of the charge in a finite interval up to arbitrary order in the interaction strength which, importantly, is capable of respecting charge quantization.

Moreover, the absence of a lower bound already creates issues in the second cumulant (that is, the local charge noise), as it gives rise to a logarithmic divergence.
The required cutoff has only been identified in noninteracting systems~\cite{Aristov1998,Ivanov2013}, while the generalization to nonzero interactions is still an unresolved fundamental problem.
As for the manifestation of charge quantization in the FCS, some recent works have explored a connection to topological phase transitions in the limited setting of transport through quantum dots~\cite{Riwar_2019b,Javed_2022}.
Furthermore, questions related to charge quantization are being discussed in circuit-quantum electron dynamics (cQED) to this day in various contexts~\cite{Likharev_1985,Loss_1991,Koch_2007,Koch_2009,Manucharyan_2009,Mizel_2020,Thanh_2020,Murani_2020,Hakonen_2021,Murani_2021,Riwar_2022,Kenawy_2022,Koliofoti_2022}.
But apart from ad-hoc recipes to ``re-quantize'' charge~\cite{Aristov1998,Mirlin2010}, such questions have barely been  addressed for 1D interacting electron systems.
We therefore believe that it is time to work towards solid-state quantum field theories capable of describing charge measurements of arbitrary spatial resolution.

Apart from the fact that the standard bosonization technique is ill-equipped to compute the charge statistics, another factor likely delayed progress.
Namely, the calculation of the generating function is a nontrivial and highly technical task already for the non-interacting case.
Starting from the well-defined formulation of the problem on the discrete 1D chain, one can map the problem to the calculation of the determinant of a large Toeplitz matrix~\cite{Klich2002,Mattis1965,Levitov1996}.
Then, the infinite-size limit of this matrix can be taken by invoking the Szeg\H{o} theorem~\cite{Grenander1958}.
However, already within the framework of the strong-limit Szeg\H{o} theorem, it turns out that its proof requires the convergence of a certain series, which is guaranteed only for $|\lambda|<\pi/3$~\cite{Mattis1965,Luttinger1963}.
The next problem arises when considering the infinite system in the limit of zero temperature.
The issue is related to the orders of these two limits and leads to the Fisher-Hartwig conjecture, which generalizes the classical strong-limit Szeg\H{o} theorem.
As is exhaustively discussed in the mathematical literature~\cite{Grenander1958,McCoyWu1973,Basor1994,Deift2011,Abanov2011,Ivanov2013}, the two limits do not commute so the order of limits is essential.
This discrepancy can be best illustrated by calculating the second cumulant, i.e., the zero-frequency noise~\cite{Abanov2011}.

In this paper, we first reiterate the precise conditions under which charge must be regarded as quantized (Sec.~\ref{sec:quantcond}), then we introduce an accurate calculation of the interaction corrections to the non-interacting cumulants generating function, test our results against the aforementioned criteria ($2\pi$-periodicity of $\m(\lambda)$ and real value at $\lambda=\pi$) in Section~\ref{sec:diags}.
We compare our results with the ones obtained by different approaches, such as Wigner crystal approximation and DMRG technique, see Sections~\ref{sec:wigner} and~\ref{sec:comparison}, respectively.

\section{Conditions for charge quantization}
\label{sec:quantcond}

To set the stage, let us briefly outline a set of assumptions which allows us to argue that the charge in any given interval must be integer-quantized.
These arguments have been outlined already in various different formulations~\cite{Haldane_1981,Rajaraman_1982,Riwar_2021}, and are reiterated here to make our work self-contained.

Take a generic fermion field $\psi(x)$, with anticommutation relations $\{\psi(x),\psi^\dagger(y)\}=\delta(x-y)$.
We now define the charge on a given interval of length $l$ as
\begin{equation}\label{eq:definition_Q}
\Q=\int_0^l dx\, \psi^\dagger(x)\psi(x) .
\end{equation}
For simplicity, we consider a translation-invariant system, so that we can choose without loss of generality the lower bound of the integral to be at $x=0$.
In the following, we focus on electron fields and set the electron charge to $1$.
Hence, the charge operator $\Q$ is dimensionless.
Note that the following argument can be easily generalized to bosons.

The only relevant assumption we need to make is that there exists a true vacuum state, $\vert 0\rangle$, defined such that $\psi(x)\vert 0\rangle =0$ for all $x$.
We stress that by vacuum state we do not mean the Fermi sea ground state containing a finite number of electrons, but really the state containing \emph{no} electrons.
Given the existence of such a true vacuum state, we can construct a complete set of many-body states with $N$ electrons, $\vert \{x_{j}\}_{N} \rangle =\psi^\dagger (x_1)\psi^\dagger (x_2)\ldots \psi^\dagger (x_q)\vert 0 \rangle$.
By means of the anticommutation relations and the definition of the vacuum state, these states can be easily shown to be eigenstates of $\Q$ with eigenvalues
\begin{equation}
\label{eq:Q1quant}
Q = \sum_{j=1}^{N}\left[\theta(x_j)-\theta(x_j-l) \right],
\end{equation}
where $\theta(x)$ denotes the Heaviside theta function.
Due to the sharpness of the $\theta$-function, these eigenvalues are integers between $0$ and $N$ depending on whether or not the electron at position $x_j$ is inside the interval $[0,l]$ of the charge measurement.
Importantly, this proof did not require any details on the Hamiltonian, and is therefore independent of interactions and strong correlations.
It is furthermore valid in arbitrary dimensions.
Hence, while the introduction of fractionally charged excitations was without doubt a milestone in understanding the physics of certain low-dimensional systems, it should be considered an effective picture valid only when some of the above assumptions can be relaxed.

Indeed, given the above proof, we can easily identify two causes for breaking charge quantization.
One possibility is that a given charge detector fails to measure the charge so precisely as to locate it with perfect certainty inside the interval $x\in[0,l]$.
Such a fuzzy detector can be modelled by a more general support function $S(x)$.
In this case, the charge operator becomes $\Q=\int dx S(x) \psi^\dagger (x) \psi (x)$ and it can have non-integer eigenvalues.
This is in agreement with the arguments put forward in Refs.~\cite{Rajaraman_1982,Riwar_2021}.

The other possibility for the above proof to fail is that there exists no true vacuum state $\vert 0 \rangle$.
The standard bosonization procedure requires removing the lower bound in the Hilbert space and continues the filled electron levels to infinite negative energies.
In the Luttinger liquid context, the two conditions are therefore related: as remarked by Haldane~\cite{Haldane_1981}, the bosonized charge density is an approximation, neglecting charge fluctuations on the length scale of $\lambda_F$ (see also the remark about backscattering in Sec.~\ref{sec:comparison}).
In fact, the genesis of Luttinger liquid theory was initially plagued exactly by these field theoretical subtleties arising from the removal of said lower bound~\cite{Luttinger1963,Mattis1965}.
However, at least if we start our field theoretic considerations from a non-relativistic standpoint, there must always exist a lower bound in the Hilbert space, and its removal is an approximation.
We can think of this procedure as a more strict version of a low-energy approximation: it is not only important, that the state of the system prior to the measurement is at low energy, but also that a given charge detection event does not give rise to high energy excitations.

Therefore, in order to answer the questions outlined in the introduction, a field theoretic treatment including the lower bound and capable of dealing with many-body correlations would be necessary.
For a generic interaction potential, the most straightforward choice is a perturbation theory in the interaction strength.
As we show in the following, when attempting to compute the moment (or cumulant) generating function, even the perturbative approach becomes rather involved.
While we can formally derive a perturbative expansion up to arbitrary order, we are nonetheless limited to the lowest orders for explicit calculations.
Moreover, in order to make progress towards the strongly interacting regime, we resort to the Wigner crystal approach, which conserves charge quantization and likewise provides a cutoff for the charge noise.
Curiously, we find that the Wigner crystal is in good qualitative agreement with the perturbative approach even for weak interactions, where it is commonly not expected to work.

\section{Diagrammatic approach to charge counting}
\label{sec:diags}

Let us begin by describing the perturbative approach.
For convenience, we consider a discrete model.
The Hamiltonian then consists of the single-particle part and the interaction, $\mathcal{H}=\mathcal{H}^{(0)}+\mathcal{V}$, where
\begin{equation}
\mathcal{H}^{(0)} = \sum_{nm} H^{(0)}_{nm}c^\dagger_{n}c_{m}, \,\,
\mathcal{V} = \frac12\sum_{nm} V_{nm} : c^\dagger_{n}c^\dagger_{m}c_{m}c_{n} :
\label{eq:hamiltonian}
\end{equation}
where the indices $m, n \in \mathbb{Z}$ run over all sites of the 1D chain, while the colon denotes normal ordering with respect to the Fermi sea.
The single-particle Hamiltonian describes the hopping and chemical potential, $H^{(0)}_{nm} = -(\delta_{n,m+1}+\delta_{n,m-1})/2 - \delta_{n,m}\sin(k_{F})$, where $k_{F}$ is the Fermi momentum and we chose the energy units such that the hopping matrix element between nearest neighbors is one.
Hence, the dimensionless parameter $k_F$ is related to the Fermi momentum of the continuum model by $p_{F}=k_{F}\delta x$, where $\delta x$ is the distance between neighbor sites.
As outlined above, our challenge is to study the counting statistics of the charge operator on an interval with $L$ sites, where $l = L \delta x$.
The charge operator for the discrete model is
\begin{equation}
\Q = \sum_{nm} Q_{nm}c^\dagger_{n}c_{m}, \label{eq:qdefinition}  
\quad
Q_{nm} = \begin{cases}
\delta_{nm} & \text{for } 1 \leq n \leq L, \\
0 & \text{otherwise}.
\end{cases} \!\!\! 
\end{equation}
The moment generating function we want to calculate can be disassembled as (see Appendix~\ref{apx:cgfdisassemle} for details)
\begin{equation}
\m(\lambda)=\langle e^{i\lambda\Q} \rangle \equiv
 \langle e^{i\lambda\Q}\rangle^{(0)} \, \langle \mathcal{S}(\beta) \rangle^{(\lambda)} \,  e^{\beta (\Omega-\Omega^{(0)})},
\label{eq:momgenfun}
\end{equation}
where
\begin{align}
\mathcal{S}(\tau) \equiv e^{\tau \mathcal{H}^{(0)}}e^{-\tau (\mathcal{H}^{(0)}+\mathcal{V})} = T_{\tau} e^{-\int_{0}^{\tau}\mathcal{V}(\tau')d\tau'}
\end{align}
denotes the (Matsubara) imaginary-time evolution operator and the operators are in the interaction picture, i.e., $\mathcal{V}(\tau) \equiv e^{\mathcal{H}^{(0)}\tau} \mathcal{V} e^{-\mathcal{H}^{(0)}\tau}$.
Moreover, $\Omega^{(0)}$ is the grand-canonical potential for the noninteracting case, such that $e^{-\beta (\Omega-\Omega^{(0)})}=\langle \mathcal{S}(\beta) \rangle^{(0)}$~\cite{AGD1965,Mahan1993}.
Different brackets are used to distinguish the averaging over the full Hamiltonian $\mathcal{H}$ [see Eq.~\eqref{eq:hamiltonian}] from the averaging over the bare Hamiltonian with a counting operator as a weight function:
\begin{equation}
\!\!
\langle \ldots \rangle
\!\equiv\! \frac{
\Tr [e^{-\beta \mathcal{H}}\! \ldots ]
}{\Tr [e^{-\beta \mathcal{H}} ]}, 
\quad
\langle \ldots \rangle^{(\lambda)}
\!\equiv\! \frac{
\Tr [e^{-\beta \mathcal{H}^{(0)}} \!\!\!\ldots e^{i\lambda \Q} ]
}{\Tr [e^{-\beta \mathcal{H}^{(0)}}  e^{i\lambda \Q} ]}.
\label{eq:average}
\end{equation}
Note that $\langle \ldots \rangle^{(0)}$ means a conventional averaging over the bare Hamiltonian as, e.g., in Ref.~\cite{AGD1965}.

\begin{figure}
\centering
\makebox[0pt]{\hspace{2cm}(a)}%
\includegraphics[width=\columnwidth]{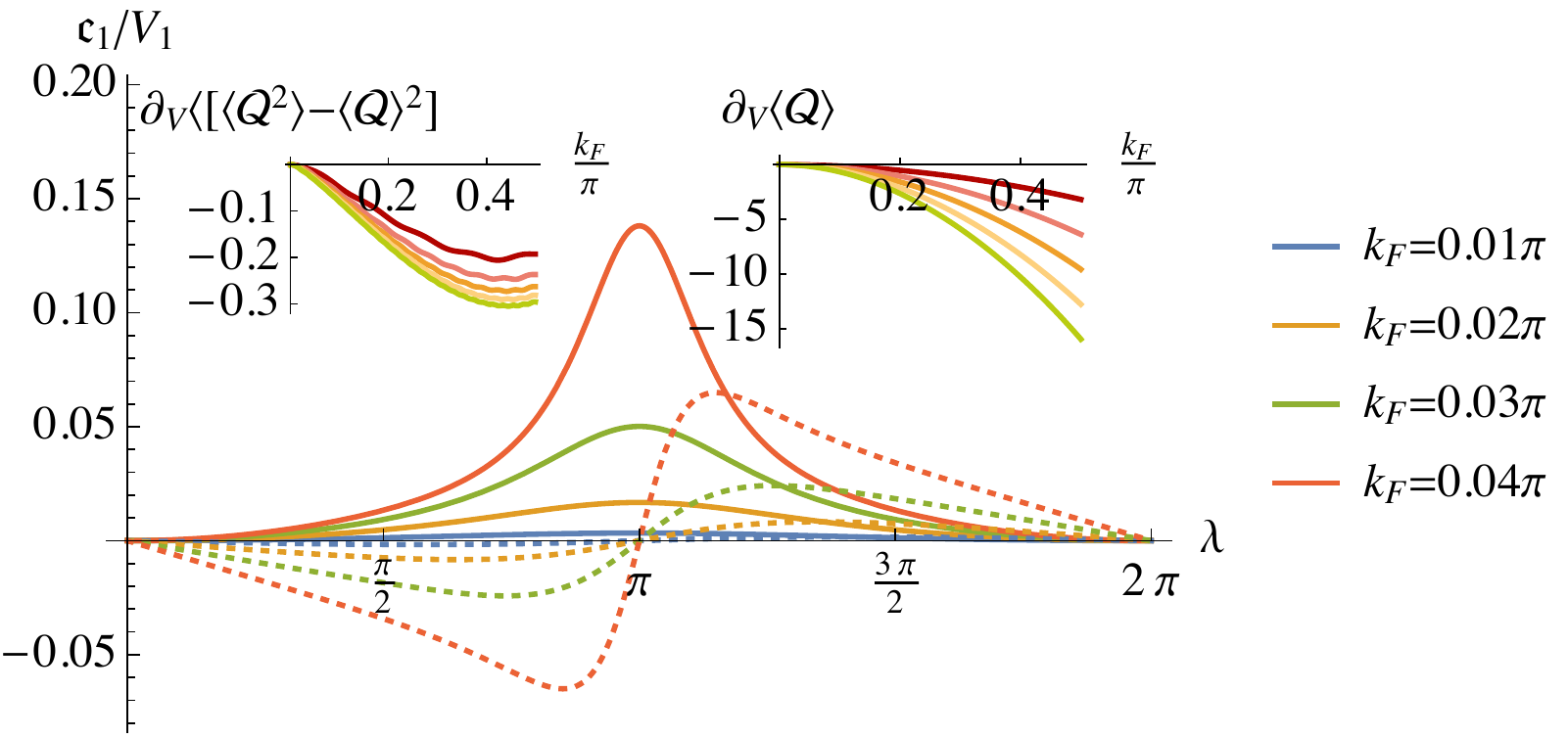}\\
\makebox[0pt]{\hspace{2cm}(b)}%
\includegraphics[width=\columnwidth]{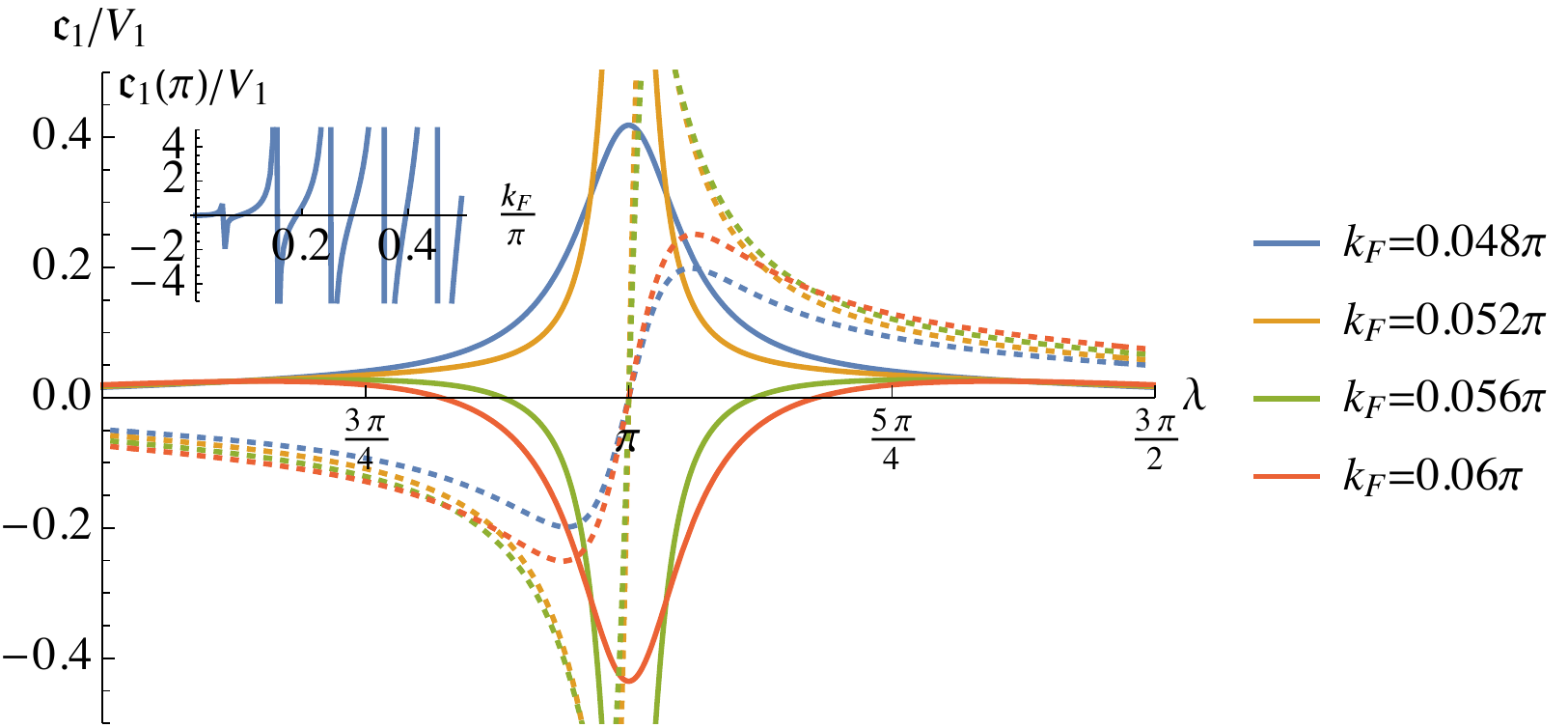}\\
\caption{%
Dependence of Hartree-Fock correction of the cumulant generating function on $\lambda$ and the occupancy for an interval of length $L=10$.
Solid and dashed lines correspond to real and imaginary components.
(a) The interaction correction to the generating function for low electron densities.
The insets demonstrate the dependence of correction to the first (right) and second (left) cumulants (i.e., noise and charge) on occupancy and show the increase of their absolute values at increase of interval length $L=10,20,30,40,50$.
The equidistance of the first cumulant lines show us an expected linear dependence on $L$.
(b) The interaction correction to the generating function for densities close to the first critical density $k_{C,1}/\pi\approx 0.054$ for $L=10$.
The inset demonstrates the sign change (or potential discontinuities) around $k_{C,n}$ of the generating function at $\lambda=\pi$.}
\label{fig:genefun}
\end{figure}

The central point of this generalized perturbation approach is that Wick's theorem is applicable for the generalized average $\left<\ldots\right>^{(\lambda)}$ as well, see Appendix~\ref{apx:wick}.
While this approach works well in general, it breaks down for certain values of the Fermi momentum at $\lambda =\pi$.
The reason for that is that the denominator $\Tr [e^{-\beta \mathcal{H}^{(0)}} (-1)^{\Q} ]$ can vanish at $k_{F}=k_{C,n} \equiv (n-\tfrac{1}{2})\pi/L$ if $L\gg 1$, see the discussion at the end of this section, the inset in Fig.~\ref{fig:genefun}(b), and Appendix~\ref{apx:convergence}.
We note that this behavior is somewhat reminiscent of the different but related context of out-of-equilibrium quantum transport, where non-analytic behavior of the FCS at $\lambda=\pi$ is routinely found when the system undergoes dissipative dynamic phase transitions, see Refs.~\cite{Ren_2012,Li_2013,Flindt_2013,Hickey_2014,Brandner_2017,Riwar_2019b,Javed_2022}.
We however believe that in this particular case, this is a spurious result, as a comparison to DMRG computations reveals that in these particular points $k_{C,n}$, the $\m(\pi)$ vanishes, i.e., the total correction $\langle \mathcal{S}(\beta) \rangle^{(\lambda)}$ is finite,  see Appendix~\ref{apx:convergence}.


The expression $\langle \mathcal{S}(\beta) \rangle^{(\lambda)}$ is identical to the expression for the thermodynamic potential up to a replacement of all Green's function by dressed Green's functions.
Since the Wick theorem works for both expressions, the graphical representation of the diagrammatic expansion will be the same, only the expressions of the basic graphical elements will differ.
Thus, expanding order-by-order in the interaction strength, the moment and cumulant generating functions ($\m(\lambda)$ and $\cm(\lambda)=\ln\m(\lambda)$, correspondingly) can be formally connected as follows
\begin{equation}
\langle e^{i\lambda\Q} \rangle
= e^{\cm_{0} + \cm_{1} + \cm_{2} + \ldots}\ ,
\label{eq:momgenfun_cum}
\end{equation}
where $\cm_{0}$ is a non-interacting result and every other term can be described by means of Feynman diagrams,
\begin{equation}
\cm_{M} =
\frac{(-1)^{M}}{M}\sum_{\substack{\text{unique} \\ \text{diagrams}}} \left(
\Bigl<\mathcal{V}^{M}\Bigr>^{(\lambda)}
-
\Bigl<\mathcal{V}^{M}\Bigr>^{(0)}
\right)
\label{eq:KMdiff} \ .
\end{equation}
As we outline in Appendix~\ref{apx:diagrams},
the connected bubble unique diagrams $\langle \mathcal{V}^{M}\rangle^{(\lambda)}$ are graphically identical to the diagrammatic expansion terms $\langle \mathcal{V}^{M}\rangle^{(0)}$ of the thermodynamic potential, the lowest-order diagrams of which are
\begin{subequations}
\label{eq:diags}
\begin{align}
\sum_{\substack{\text{unique} \\ \text{diagrams}}}\Bigl<\mathcal{V}^{1}\Bigr>^{(0)} = & \raisebox{-0.14in}{\includegraphics[scale=0.7,page=1]{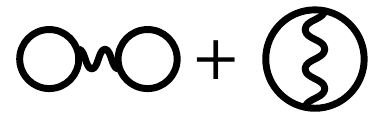}}
\label{eq:diags1hf}
\\
\sum_{\substack{\text{unique} \\ \text{diagrams}}}\Bigl<\mathcal{V}^{2}\Bigr>^{(0)} = & \raisebox{-0.14in}{\includegraphics[scale=0.7,page=2]{diags}} \nonumber\\
& \raisebox{-0.14in}{\includegraphics[scale=0.7,page=3]{diags}} \ .
\label{eq:diags2}
\end{align}
\end{subequations}
Our main result is that the nontrivial
second term in Eq.~\eqref{eq:momgenfun} corresponds merely to \emph{interpreting} the standard diagrams of a known perturbation series in terms of dressed Green's functions $G^{(\lambda)}$, which are merely a conventional Matsubara Green's functions $G^{(0)}$ dressed with the counting operators:
\begin{equation}
G^{(\lambda)}_{nm}(\tau_{1},\tau_{2}) \equiv
-\left<T_{\tau} c_{n}(\tau_{1})c_{m}^{\dagger}(\tau_{2})\right>^{(\lambda)}
\label{eq:MGF}
\end{equation}
Note that with this definition the zero Green's function $G^{(0)}(\tau_{1},\tau_{2})=G^{(0)}(\tau_{1}-\tau_{2})$ is the conventional Matsubara Green's function
\begin{equation}
G^{(0)}(\tau) = e^{-H^{(0)}\tau}\left[\left(e^{\beta H^{(0)}}+1\right)^{-1}-\theta(\tau)\right] \ .
\label{eq:GF0}
\end{equation}
The generalized and conventional Green's functions can be easily related using the parameter $\zeta= e^{i\lambda}-1$, which thus becomes the only way in which the $\lambda$-dependence enters the equation.
This also makes it clear that all dressed Green's functions are $2\pi$-periodic in $\lambda$ and have zero imaginary part at $\lambda=\pi$, as required by charge quantization.
The explicit relation is
\begin{equation}
G^{(\lambda)}(\tau_{1},\tau_{2}) = G^{(0)}(\tau_{1}-\tau_{2}) + \tilde{G}^{(\lambda)}(\tau_{1},\tau_{2}),
\label{eq:GFl}
\end{equation}
where
\begin{equation}
\begin{split}
\tilde{G}^{(\lambda)}(\tau_{1},\tau_{2}) &= - \zeta G^{(0)}(\tau_{1}) QD^{-1}Q  G^{(0)}(-\tau_{2}), \\
D &= 1 + \zeta QG^{(0)}(-0)Q.
\end{split}
\label{eq:GFlb}
\end{equation}
One can account for the projection operator $Q$ by a reduced summation over the interval $i,j \in [1, L]$ where the charge is measured, see Eq.~\eqref{eq:qdefinition}.
Thus, in Eq.~\eqref{eq:GFlb} we obtain $\tilde{G}^{(\lambda)}_{nm}(\tau_1,\tau_2) = - \zeta \sum_{ij} G^{(0)}_{ni}(\tau_1) (D^{-1})_{ij}  G^{(0)}_{jm}(\tau_2)$, where the matrix $D$ can be treated as a matrix of size $L\times L$ with elements $D_{ij}=\delta_{ij}+\zeta G^{(0)}_{ij}(-0)$.
In both cases the non-interacting part of the generating function can be written in the form of a Fredholm determinant $\cm_{0}=\ln\det D = \Tr \ln D$~\cite{Klich2002,Levitov1996}.

The first order correction due to interactions, $\cm_{1}$, is described by the Hartree-Fock terms illustrated in Eq.~\eqref{eq:diags1hf}.
Splitting the dressed Green's functions as shown in Eq.~\eqref{eq:GFl} we obtain from Eq.~\eqref{eq:KMdiff}
\begin{multline}
\cm_{1} = -\frac{1}{2}\sum_{ij} V_{ij}
\int_{0}^{\beta} d\tau
\Bigl\{
2G^{(0)}_{ii}\tilde{G}^{(\lambda)}_{jj}
-
2G^{(0)}_{ij}\tilde{G}^{(\lambda)}_{ji}
+\\+
\tilde{G}^{(\lambda)}_{ii}\tilde{G}^{(\lambda)}_{jj}
-
\tilde{G}^{(\lambda)}_{ij}\tilde{G}^{(\lambda)}_{ji}
\Bigr\},
\label{eq:K1def}
\end{multline}
where all Green's functions depend on $\tau$ as $G=G(\tau,\tau)$.
The explicit expression can be found in Appendix~\ref{apx:HF}.
Assuming a translation-invariant interaction potential $V_{ij}=V_{|i-j|}$, the only non-zero elements for the case of nearest-neighbor interactions are $V_{0}$ and $V_1$.
Note that $V_0$ is irrelevant because we consider spinless fermions for which the Pauli principle rules out double occupation of a given site.
For low filling we present in Fig.~\ref{fig:genefun}(a) the numerical result demonstrating a $2\pi$-periodic $\lambda$-dependence of the cumulant generating function.
One can observe that the value of $\cm_{1}$ first increases with occupation.
Moreover, the first and second cumulants, i.e., the charge and noise, keeps doing so practically at all filling factors (see insets).
One can also see that the charge expectation value is linear with the interval length $L$ (the lines are equidistant, see caption).
The analysis of the noise dependence on the occupation (shown in left inset of Fig.~\ref{fig:genefun}a) and other system parameters is more sophisticated and is done in the Section~\ref{sec:comparison}, which is devoted to this problem.
The above mentioned growth with the occupation, which is always correct for the low cumulants and valid for small occupancies of generating function, however, is not true for finite $\lambda$, especially close to $\lambda=\pi$.
The generating function is indeed real at this point for low occupations, as well as close to the first critical densities $k_{C,n}$, see Fig.~\ref{fig:genefun}b.
We can claim that the value $\cm_{1}(\lambda=\pi)$ indeed changes sign when passing these critical occupations, but the lowest order of perturbation theory is not sufficient to find out the exact behavior around $k_{C,n}$ points, so it remains unclear whether it is a smooth crossover or a discontinuity.
Nevertheless, the DMRG studies demonstrate that the total interaction correction to the cumulant generating function $\cm-\cm_{0}$ remains finite, so the behavior of the $\mathfrak{m}(\pi)$ is governed by $e^{\cm_{0}}$ keeping the zeros at $k_{C,n}$ unchanged (see Appendix~\ref{apx:convergence} for details).

\section{Wigner crystal}
\label{sec:wigner}

While we have succeeded in formulating the FCS of interacting electrons in terms of a diagrammatic approach up to in principle arbitrary order, this approach is for practical purposes obviously limited to lowest order contributions, and thus does not allow us to venture into the regime of strong correlations.
For this purpose, we have to find some other means.
As it turns out, another viable strategy is to compute the FCS for the situation where the electrons form a Wigner crystal.
Of course it can be expected that this picture provides reliable results for strong repulsive interactions.
But with the previous method, we have the unique opportunity to test, whether this picture might work also for weak interactions.


The Wigner crystal can be seen as a discretized version of the Luttinger liquid~\cite{Matveev2007,Meyer_2009} in a semiclassical regime: the spatial variation of the bosonic field is dominantly realized by kinks (see Appendix~\ref{apx:wigner}), such that the kink position can be directly related to the localization of an electron charge.
The dynamics of these kinks can be described in terms of a chain of $N$ serially connected harmonic oscillators.
Despite of the bosonic nature of such system, its validity is not as far-fetched as it may initially seem: strong repulsive interaction prevents the violation of the Pauli exclusion expected from the electrons.
In the subsequent section, we demonstrate qualitative agreement with perturbative results for the fermionic system, indicative of the fact that repulsion by the Pauli principle itself (in the absence of strong interactions) is well-approximated by the oscillator chain, too.
In order to control the electron density and prevent the collapsing the electrons, it is convenient to place the oscillator chain on a ring to ensure the stability of the system,
\begin{equation}
H = \sum_{n=1}^{N} \left[\frac{p_{n}^{2}}{2m} + \frac{m\omega^2}{2}(x_{n}-x_{n+1})^{2} \right]
\label{eq:oscillatorschain}
\end{equation}
with the canonically conjugate oscillator positions and momenta, $[x_k,p_{k'}]=i\delta_{kk'}$.
The periodic boundaries are imposed such that $x_{N+1}\equiv x_{1}+N/\varrho$, where $N/\varrho$ is a circumference of a ring, $\varrho$ is the density of oscillators corresponding to the density of original electronic excitations $\varrho=p_{F}/\pi$.
We keep using the notation $\varrho$ to distinguish the oscillator chain from the actual fermionic model.
The oscillator parameters are given such that $m\omega=\pi \varrho^2 /K$.
We see immediately, that the oscillator parameters on the one hand connect seamlessly to the Luttinger liquid interaction parameter $K$, and at the same time, the system knows about the total electron density $\varrho$.
Consequently, the average charge (number of oscillators) on the interval of the length $l$ is equal to $\langle \Q \rangle = \varrho l$.

The charge inside the interval $[0,l]$ is computed by testing whether a given oscillator is in it.
Thus, the expectation value of $\Q^M$ for the Wigner crystal can be written as
\begin{equation}
\langle \Q^{M} \rangle =  \int dx_1\!\ldots\!\int dx_N  Q^{M} \vert\psi_0(x_1,\ldots,x_N)\vert^2\ ,
\end{equation}
where $Q$ is given by Eq.~\eqref{eq:Q1quant} and $\psi_0(x_1,\ldots,x_N)$ is the wave function of the ground state of Eq.~\eqref{eq:oscillatorschain}.
%
As a consequence, the moment generating function is here likewise by construction $2\pi$-periodic in $\lambda$.
In addition, the theory has a natural cut-off for the second cumulant, due to the granular nature of the charge density.
In particular, the prediction for the second moment in such Wigner crystal is (for details on the calculation, see Appendix~\ref{apx:wigner}).
\begin{equation}
\langle \Q^{2} \rangle \!=\! \varrho l \!+\! \sum_{j=1}^{\infty} \! b_{j} \!
\left[
\mathrm{erfi}\frac{\varrho l\!-\!j}{b_{j}} \!+\! \mathrm{erfi}\frac{\varrho l\!+\!j}{b_{j}} \!-\! 2\mathrm{erfi}\frac{j}{b_{j}}
\right],
\label{eq:wigner}
\end{equation}
where $\mathrm{erfi}\,x = \pi^{-1/2}(e^{-x^{2}}+2x\int_{0}^{x}e^{-y^{2}}dy)$ and $b_{j} = \frac{2}{\pi}\sqrt{K(1+\frac12\ln j)}$.
The noise, i.e., the fluctuation of the number of oscillators in the interval, can be obtained through $\langle \Q^{2} \rangle - \langle \Q \rangle^{2}$.

\section{Comparison}
\label{sec:comparison}

In order to compare the outcome of the above approaches we choose as a reference object the second cumulant, i.e., the zero-frequency charge noise, described by the well-known formula $\pi^{-2}\ln(\kappa L)$
for the non-interacting case~\cite{LeHur2012,Aristov1998}.
This comparison will also help us to resolve the discrepancy in the exact expression for the logarithmic dependence cutoff $\kappa$ that one can find in literature.
In the non-interacting case the value of the cutoff for the interval size $L$ was calculated by means of the so called strong Szeg\H{o} theorem~\cite{Luttinger1963,Aristov1998,McCoyWu1973}: $2e^{\gamma_\text{E}}\sin k_{F}$, where $\gamma = \gamma_{E}\approx 0.5772$ is Euler's constant (the digest of the calculation is given the Eqs.~(21)--(25) in Ref.~\cite{Aristov1998}).
This result rests on a calculation at finite temperature and taking the limit $T\to0$ afterwards.
The accurate calculation in the case of setting $T=0$ from the start with subsequent limit $L\gg1$ requires taking Fisher-Hartwig singularities into account~\cite{Basor1994,Deift2011,Abanov2011,Ivanov2013}.
This gives a slightly different answer $2e^{\gamma_\text{E}+1}\sin k_{F}$ (see detailed calculation in Appendix~B.2 of Ref.~\cite{Abanov2011}), which is also proved by direct noise calculations~\cite{LeHur2012}.


As already pointed out in the introduction, the situation is even more sophisticated if interactions are added.
While bosonization proved to be an extremely powerful tool for describing the interacting fermion systems, due to the lack of a lower bound, it is unable to correctly account for the cutoff in the logarithm.
Moreover, within the bosonic representation, the result for the generating function contains only $1^\text{st}$ and $2^\text{nd}$ cumulants~\cite{Aristov1998}, which immediately breaks $2\pi$-periodicity.
The requirement to be real at $\lambda=\pi$ is broken as well, and there are arguments that the missing $2p_{F}$  backscattering processes are responsible for this discrepancy~\cite{Aristov1998,Luther1975}.
However, there is an open question why these processes are less relevant for other values of $\lambda$.

\begin{figure}
\centering
\includegraphics[width=\columnwidth]{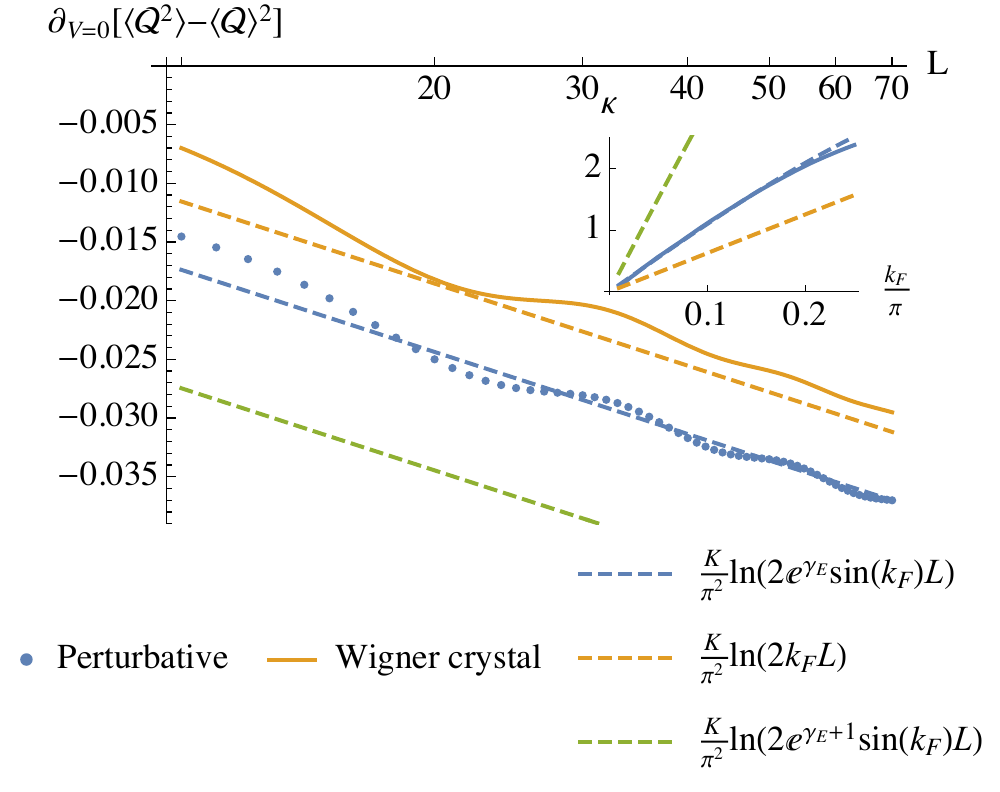}
\caption{The noise lowest order interaction correction and its dependence on occupancy $k_{F}$ and the length of the measuring interval $L$.
The main logarithmic plot demonstrates the dependence on $L$ at $k_{F}=\pi/20$.
Dots correspond to our perturbative approach, solid curve is Wigner crystal result, and dashed lines are interaction corrections obtained from the formulas given in the plot's legend.
The offset of the data in the main plot is governed by the cutoff parameter $\kappa$ [see Eqs.~\eqref{eq:kappanoise} and~\eqref{eq:kappanoiseintcorr}] pictured in the inset, which color scheme labeling corresponds to the legend of the main plot.
}
\label{fig:noise}
\end{figure}

We compare the results for the noise obtained by all three approaches. %
All three results are illustrated in Fig.~\ref{fig:noise}.
i) Using bosonization methods as in Ref.~\cite{Aristov1998}, one may merely estimate
\begin{equation}
\langle \Q^{2} \rangle - \langle \Q \rangle^{2} = \frac{K}{\pi^{2}}\ln(\kappa L)
\label{eq:kappanoise}
\end{equation}
where the cutoff $\kappa$ is added by hand, and usually chosen to be of order of Fermi momentum $\sim k_{F}$.
ii) The expansion of the Eq.~\eqref{eq:K1def} in orders of $\lambda$ and approximate calculation for $k_{F}\ll 1$ of the integrals gives 
\begin{equation}
\partial_{\lambda=0}^{2}\cm_{1}\approx\frac{2V_{1}\sin k_{F}}{\pi^{3}}\ln(2k_{F}L).
\label{eq:kappanoiseintcorr}
\end{equation}
iii) The result of the Wigner crystal model was already given by Eq.~\eqref{eq:wigner}.

\begin{figure}
\centering
\includegraphics[width=.95\columnwidth]{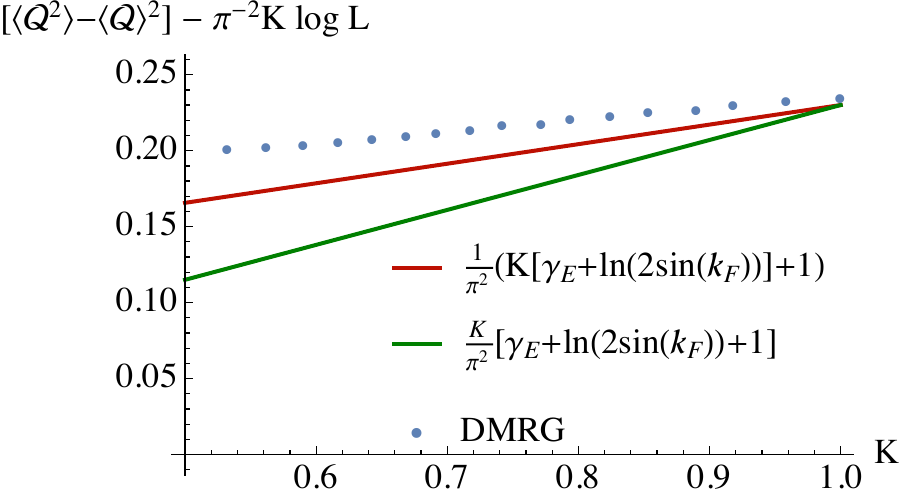}
\caption{Interaction dependence of the offset.
Dots are obtained by DMRG simulation, while lines illustrate the formulas in the the legend at the occupancy of the $k_{F}=\pi/2$.}
\label{fig:gdmgr}
\end{figure}

Let us first discuss the relationship between i) and ii). If we choose the cutoff  $\kappa=2k_{F}$ in Eq.~\eqref{eq:kappanoise}, the two results agree, since the Luttinger parameter for weak interaction is $K = 1 - \frac{2}{\pi}V_{1}\sin k_{F}$.
We further note that we can increase the precision, and compute the integrals in Eq.~\eqref{eq:K1def} numerically. We thus uncover a refined value for the cutoff of the asymptotic logarithmic behavior, see the inset of Fig.~\ref{fig:noise}.
%
One may see, that the numerical calculation of the cutoff parameter in the interacting part (blue dots) is extremely good agrees with the non-interacting formula from Ref.~\cite{Luttinger1963}, with $\gamma_{E}$ only (blue dashes) multiplied by $1-K$.
The accurate for zero temperature case non-interacting cutoff from Ref.~\cite{Abanov2011}, with $\gamma_{E}+1$ (green dashes), meanwhile, does not fit that well.
Combining the noninteracting result~\cite{Luttinger1963} with ours for the limit of small $1-K$, we get an exact formula for the noise (contrary to already existing results in the literature where the cutoff is only estimated)
\begin{equation}
\langle \Q^{2} \rangle - \langle \Q \rangle^{2} = \frac{1}{\pi^{2}}[K(\ln(2L\sin k_{F}) + \gamma_\text{E})+1]\ .
\label{eq:noisep1}
\end{equation}
Our result allows to interpolate from the noninteracting case into the interacting regime.
%
In order to further corroborate the logarithmic cutoff result (including the $\gamma_{E}$ versus~$\gamma_{E}+1$ issue), we resorted to a DMRG approach, which allows to analyze the system at wide range of interaction strength (from $K=0$ to $K=1/2$ in our case).
The numerical DMRG results for the periodic boundary conditions are illustrated in Fig.~\ref{fig:gdmgr} demonstrating the substantially better fit of the numerical data with Eq.~\eqref{eq:noisep1} in a very broad interval of the interaction strength.

Finally, let us discuss the Wigner crystal result iii), in the regime of weak interactions. We note that while it does not reproduce the cutoff with the same numerical accuracy as Eq.~\eqref{eq:noisep1}, it nonetheless correctly captures quite a number of qualitative features, such as the correct decrease of charge noise with the onset of repulsive interactions (i.e., all results in Fig.~\ref{fig:noise} are negative). Moreover, it neatly reproduces the oscillations of the noise as a function of $L$, which can also be seen in the perturbative results. Note in particular, that the period of the oscillations, going with $\sim\!\kappa^{-1}$, match perfectly between the perturbative and Wigner crystal approach. Such oscillations cannot possibly be reproduced by i), which is rooted in the very nature of standard bosonization.
These observations speak in favor of a high  validity of the Wigner crystal approximation even for weak interactions, being able to mimic Pauli's exclusion principle by means of a primitive oscillators chain.

\section{Conclusions}

This work contains several important results regarding the full counting statistics in one-dimensional systems of interacting fermions.
First, we developed an universal perturbative approach that allows us to calculate the interaction corrections to the cumulant generating function for any value 
(with an exception of value $\pi$ at particular electron densities only) 
of the counting field $\lambda$, preserving the real value of generating function at $\lambda=\pi$ and its $2\pi$-periodicity.
%
%
%
%
%
%
Second, using this approach, we calculated the accurate expression for the noise for the interacting case and determined an exact value for the logarithmic cutoff at zero and finite interaction.
By means of these accomplishments, we showed that the Wigner crystal approach may be used for good qualitative predictions of charge counting even in the case of weak interactions.

\begin{acknowledgments}
This work has been funded by the German Federal Ministry of Education and Research within the funding program Photonic Research Germany under the contract number 13N14891. TLS and AH acknowledge financial support from the National Research Fund Luxembourg under grants CORE C20/MS/14764976/TopRel and INTER/17549827/AndMTI.
\end{acknowledgments}

\appendix

\setcounter{figure}{0}
\renewcommand{\thefigure}{\Alph{section}\arabic{figure}}




%
%

\section{Generating function disassembling}
\label{apx:cgfdisassemle}
\setcounter{figure}{0}

The moment generating function we need to calculate is defined and can be rewritten in the following way:
\begin{align}
\langle e^{i\lambda\Q} \rangle\equiv &
\frac{\Tr \left[e^{-\beta (\mathcal{H}^{(0)} + \mathcal{V})} e^{i\lambda \Q}\right]}{\Tr \left[e^{-\beta (\mathcal{H}^{(0)} + \mathcal{V})}\right]}
\nonumber\\ = &
\frac{\Tr \left[e^{-\beta (\mathcal{H}^{(0)} + \mathcal{V})} e^{i\lambda \Q} \right]}{\Tr \left[ e^{-\beta \mathcal{H}^{(0)}}e^{i\lambda \Q}\right]}
\times\nonumber\\\times&
\frac{\Tr \left[ e^{-\beta \mathcal{H}^{(0)}}e^{i\lambda \Q}\right]}{\Tr \left[e^{-\beta \mathcal{H}^{(0)} }\right]}
\left[\!\frac{\Tr \left[e^{-\beta (\mathcal{H}^{(0)} + \mathcal{V})}\right]}{\Tr \left[e^{-\beta \mathcal{H}^{(0)}}\right]}\!\right]^{-1}\!\!.
\label{eq:countingsplit}
\end{align}
The individuals terms in this expression have the following interpretations:
\begin{itemize}
\item The \emph{third} term is a correction to the thermodynamical potential, which can be expressed in a conventional series of Feynman diagrams, see Eq.~\eqref{eq:diagrams}, consisting of conventional Matsubara Green's functions $G^{(0)}$ given in Eq.~\eqref{eq:GF0}.
\item The \emph{second} term is the counting operator of the noninteracting system, $\langle e^{i\lambda \Q}\rangle^{(0)}=e^{\cm_{0}}$, since the trace are taken with respect to the single-particle Hamiltonian $\mathcal{H}^{(0)}$ only.
\item The \emph{first} term can be expanded in the interaction Hamiltonian $\mathcal{V}$ in the same way as the thermodynamic potential in Eq.~\eqref{eq:diagrams}, forming averages $\langle\mathcal{V}^{M}e^{i\lambda \Q}\rangle^{(0)}$.
These averages are nothing but the expressions in Eq.~\eqref{eq:caverage}, which can be split into the pairwise averages according to the generalized Wick theorem in Appendix~\ref{apx:wick} and in particular Eq.~\eqref{eq:caveragewick}.
\end{itemize}

Thus, the \emph{first} term can be calculated by building a diagrammatic expansion using Green's functions dressed with a counting operator $e^{i\lambda \Q}$.
This dressed Matsubara Green's function is defined in Eq.~\eqref{eq:MGF}.
Taking the expression for the thermodynamic potential and using it in Eq.~\eqref{eq:countingsplit}, we obtain the diagrammatic expansion
\begin{multline}
\ln\langle e^{i\lambda\Q} \rangle - \ln\langle e^{i\lambda\Q} \rangle^{(0)} =
\sum_{M=1}^{\infty}\frac{(-1)^{M}}{M}
\times\\\times\sum_{\substack{\text{all topologically} \\ \text{unique expressions}}} \left(
\stackrel{\text{unique}}{\Bigl<\mathcal{V}^{M}\Bigr>^{(\lambda)}}
-
\stackrel{\text{unique}}{\Bigl<\mathcal{V}^{M}\Bigr>^{(0)}}
\right),
\end{multline}
where in the diagrams of the second term $\langle\mathcal{V}^{M}\rangle^{(0)}_{\text{unique}}$, all lines correspond to bare Green's function $G^{(0)}$ while in the first term $\langle\mathcal{V}^{M}\rangle^{(\lambda)}_{\text{unique}}$ all lines are dressed Green's function $G^{(\lambda)}$.

\section{Wick's theorem for dressed Green's functions}
\label{apx:wick}
\setcounter{figure}{0}

The conventional Wick theorem states that an average of a product of ladder operators can be written as a sum over all possible pairings of operator averages.
In this appendix, we will show that the theorem remains true also for the dressed Green's function.
Then, we can express Wick's theorem as
\begin{align}
\langle c_{1} c_{2}^{\dagger}\ldots c_{2n-1} c_{2n}^{\dagger}\rangle^{(0)} &\equiv
\frac{\Tr\left[e^{-\beta\mathcal{H}_{0}}c_{1}c_{2}^{\dagger}\ldots c_{2n-1} c_{2n}^{\dagger}\right]}
{\Tr\left[e^{-\beta\mathcal{H}_{0}}\right]} \notag \\
&\hspace{-1.2cm}=
\sum_{\{i_{k}\}} \pm\langle  c_{i_{1}} c_{i_{2}}^{\dagger}\rangle^{(0)}\ldots\langle c_{i_{2n-1}} c_{i_{2n}}^{\dagger}\rangle^{(0)}\!.
\label{eq:conventionalwick}
\end{align}
where the sum is over all possible pairings and the $\pm$ sign depends on the parity of the chosen pairing.

To demonstrate Wick's theorem for dressed Green's function, we use their definition and expand the exponent as follows,
\begin{widetext}
\begin{align}
& \langle  c_{1} c_{2}^{\dagger}\ldots c_{2n-1} c_{2n}^{\dagger}e^{i\lambda\Q}\rangle^{(0)}
=
\sum_{m=0}^{\infty} \frac{(i\lambda)^{m}}{m!}\langle  c_{1} c_{2}^{\dagger}\ldots c_{2n-1} c_{2n}^{\dagger}\Q^{m}\rangle^{(0)}
\notag \\
&= \Bigl/ \substack{%
\text{the total expression is split (with corresponding combinatorial weights $C^{n}_{m}$) into the open}%
\\%
\text{lines of $\Q$'s connecting external ladder operators, and closed circles containing only $\Q$s}%
}\Bigr/= \notag\\
&=
\sum_{m=0}^{\infty} \frac{(i\lambda)^{m}}{m!}
\sum_{\{i_{k}\}} \sum_{\{m_{k}\}}^{\sum m_{k} = m}
\pm C^{m_{1}}_{m}\langle  c_{i_{1}} c_{i_{2}}^{\dagger}Q^{m_{1}}\rangle_{c}^{(0)}
C^{m_{2}}_{m-m_{1}}\langle  c_{i_{3}} c_{i_{4}}^{\dagger}Q^{m_{2}}\rangle_{c}^{(0)}
\ldots
\overbrace{C^{m_{n}}_{m_{n}+m_{n+1}}\langle c_{i_{2n-1}} c_{i_{2n}}^{\dagger}Q^{m_{n}}\rangle_{c}^{(0)}}^{\text{one open line}}
\underbrace{\langle Q^{m_{n+1}} \rangle^{(0)}}_{\text{all circles}}
\notag \\
&=\bigl/ {\footnotesize%
\text{expanding the expressions for the binomial coefficients $C^{n}_{m}$}%
}\bigr/= \notag\\
&=
\sum_{m=0}^{\infty} (i\lambda)^{m}
\sum_{\{i_{k}\}} \sum_{\{m_{k}\}}^{\sum m_{k} = m}
\pm
\frac{1}{m_{1}!}\langle  c_{i_{1}} c_{i_{2}}^{\dagger}Q^{m_{1}}\rangle_{c}^{(0)}
\frac{1}{m_{2}!}\langle  c_{i_{3}} c_{i_{4}}^{\dagger}Q^{m_{2}}\rangle_{c}^{(0)}
\ldots
\frac{1}{m_{n}!m_{n+1}!}\langle c_{i_{2n-1}} c_{i_{2n}}^{\dagger}Q^{m_{n}}\rangle_{c}^{(0)} \langle Q^{m_{n+1}} \rangle^{(0)}
\end{align}
where $m_{1}+\ldots m_{n} + m_{n+1} = m$, such that $0\leq m_{k}\leq m$ and $C_{n}^{m}=\frac{n!}{m!(n-m)!}$.
The sign is chosen in the same way as in Eq.~\eqref{eq:conventionalwick}, according to the number of permutation of fermion operators.
The subscript $c$ is used to indicate ``connected'' diagrams.
Now we apply the conventional Wick theorem to the obtained expressions.
\emph{This procedure is valid if $L\gg m$.}
The above can be rewritten with the sums over $m_{k}$ taken from zero to infinity
\begin{multline}
\sum_{\{i_{k}\}} \sum_{\{m_{k}\}=0}^{\infty}\pm \frac{(i\lambda)^{m_{1}}}{m_{1}!}\langle  c_{i_{1}} c_{i_{2}}^{\dagger}Q^{m_{1}}\rangle_{c}^{(0)}
\frac{(i\lambda)^{m_{2}}}{m_{2}!}\langle  c_{i_{3}} c_{i_{4}}^{\dagger}Q^{m_{2}}\rangle_{c}^{(0)}
\ldots 
\frac{(i\lambda)^{m_{n}}}{m_{n}!}\langle c_{i_{2n-1}} c_{i_{2n}}^{\dagger}Q^{m_{n}}\rangle_{c}^{(0)}
\frac{(i\lambda)^{m_{n+1}}}{m_{n+1}!} \langle Q^{m_{n+1}} \rangle^{(0)}
= \\ =
\sum_{\{i_{k}\}} \pm \langle  c_{i_{1}} c_{i_{2}}^{\dagger}e^{i\lambda Q}\rangle_{c}^{(0)}
\langle  c_{i_{3}} c_{i_{4}}^{\dagger}e^{i\lambda Q}\rangle_{c}^{(0)}
\ldots
\langle c_{i_{2n-1}} c_{i_{2n}}^{\dagger}e^{i\lambda Q}\rangle_{c}^{(0)}
\langle e^{i\lambda Q} \rangle^{(0)}
\end{multline}
\end{widetext}
For the particular case of two operators this formula can be written as
\begin{equation}
\langle  c_{1} c_{2}^{\dagger}e^{i\lambda\Q}\rangle^{(0)}
=
\langle  c_{1} c_{2}^{\dagger}e^{i\lambda\Q}\rangle_{c}^{(0)}
\langle e^{i\lambda\Q}\rangle^{(0)}
\end{equation}
Introducing a generalized $\lambda$-average, which so far we used in the sense ``connected''
\begin{multline}
\langle  c_{1} c_{2}^{\dagger}\ldots c_{2n-1} c_{2n}^{\dagger}\rangle^{(\lambda)}
\equiv
\frac{
\langle  c_{1} c_{2}^{\dagger}\ldots c_{2n-1} c_{2n}^{\dagger}e^{i\lambda\Q}\rangle^{(0)}}{
\langle e^{i\lambda\Q}\rangle^{(0)}
}
=\\=
\frac{
\Tr\left[e^{-\beta\mathcal{H}_{0}}c_{1}c_{2}^{\dagger}\ldots c_{2n-1} c_{2n}^{\dagger}e^{i\lambda\Q}\right]
}{
\Tr\left[e^{-\beta\mathcal{H}_{0}}e^{i\lambda\Q}\right]
}
\label{eq:caverage}
\end{multline}
so that for the particular two-operators case we get $\langle  c_{1} c_{2}^{\dagger}\rangle^{(\lambda)} \equiv \langle  c_{1} c_{2}^{\dagger}e^{i\lambda\Q}\rangle_{c}$,
one can formulate Wick theorem for the generalized $\lambda$-average:
\begin{multline}
\langle  c_{1} c_{2}^{\dagger}\ldots c_{2n-1} c_{2n}^{\dagger}\rangle^{(\lambda)}
=\\=
\sum_{\{i_{k}\}} \pm\langle  c_{i_{1}} c_{i_{2}}^{\dagger}\rangle^{(\lambda)}\ldots\langle c_{i_{2n-1}} c_{i_{2n}}^{\dagger}\rangle^{(\lambda)}
\label{eq:caveragewick}
\end{multline}
This equation constitutes the generalized Wick theorem.

\section{Thermodynamic potential and bubble diagrams}
\label{apx:diagrams}
\setcounter{figure}{0}

In this appendix we briefly repeat the known results for the perturbative expansion of the thermodynamic potential, which can be found in classical textbooks, e.g., Refs.~\cite{Mahan1993} or~\cite{AGD1965}.
Feynman diagrams are usually drawn as connected graphs with one or several incoming and outgoing lines, while the disconnected parts of the diagrams are cancelled out.
In the diagrammatic expansion of the thermodynamic potential the situation is different because one needs to calculate these disconnected diagrams themselves.
The definition of the thermodynamic potential is $\Omega=-T\ln\Tr  e^{-\beta (\mathcal{H}^{(0)}+\mathcal{V})}$ while in the absence of interactions it is $\Omega^{(0)}=-T\ln\Tr  e^{-\beta \mathcal{H}^{(0)}}$.
Repeating the standard steps of the diagrammatic calculation, namely expanding in $\mathcal{V}$ in the interaction representation and applying Wick's theorem we get
\begin{equation}
\exp\left[-\beta(\Omega-\Omega^{(0)})\right] =  1 + \sum_{M=1}^{\infty}
\frac{(-1)^{M}}{M!}
\Big\langle\mathcal{V}^{M}\Big\rangle^{(0)}
\label{eq:expansionthermpot}
\end{equation}
where the average on the right is defined in Eq.~\eqref{eq:conventionalwick} and the integrations over the internal Matsubara time variables are implied.
Summing up all disconnected diagrams, one can show that Eq.~\eqref{eq:expansionthermpot} can be rewritten as \cite{Mahan1993}
\begin{equation}
\Omega-\Omega^{(0)} =  -T \sum_{M=1}^{\infty}
\frac{(-1)^{M}}{M!}
\stackrel{\text{connected}}{\Bigl<\mathcal{V}^{M}\Bigr>_{c}^{(0)}}
\label{eq:expansionthermpotconnected}
\end{equation}
where the sum of only over connected diagrams.
However, a permutation of the interaction vertexes
As a result, the correction to the thermodynamic potential due to the interaction is equal to
\begin{equation}
\Omega-\Omega^{(0)} =
\sum_{M=1}^{\infty}\frac{(-1)^{M}}{M}\sum_{\substack{\text{all topologically} \\ \text{unique diagrams}}} \stackrel{\text{unique}}{\Bigl<\mathcal{V}^{M}\Bigr>^{(0)}}
\end{equation}
where the average on the right denotes a single term of the Wick expansion corresponding to a particular diagram.
Using the symmetrized form of the two-particle interaction, these bubble diagrams are (we provide only the first three orders)
\begin{subequations}
\label{eq:diagrams}
\begin{align}
\sum_{\substack{\text{all topologically} \\ \text{unique diagrams}}}
\stackrel{\text{unique}}{\Bigl<\mathcal{V}^{1}\Bigr>^{(0)}} = & \raisebox{-0.2in}{\includegraphics[scale=0.5,page=1]{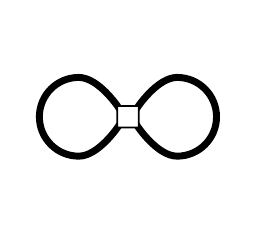}}
\\
\sum_{\substack{\text{all topologically} \\ \text{unique diagrams}}}
\stackrel{\text{unique}}{\Bigl<\mathcal{V}^{2}\Bigr>^{(0)}} = & \raisebox{-0.2in}{\includegraphics[scale=0.5,page=2]{thermopot}}
\\
\sum_{\substack{\text{all topologically} \\ \text{unique diagrams}}}
\stackrel{\text{unique}}{\Bigl<\mathcal{V}^{3}\Bigr>^{(0)}} = & \hspace{-0.25in}\raisebox{-0.7in}{\includegraphics[scale=0.5,page=3]{thermopot}}
\end{align}
\end{subequations}
In our paper is is more convenient to use the initial, not symmetrized form of the interaction, as we demonstrated for the first two lines in Eq.~\eqref{eq:diags}.

\section{Hartree-Fock}
\label{apx:HF}

The formula for the Hartree-Fock contribution pictured in Fig.~\ref{fig:genefun} can be obtained by substituting Eqs.~\eqref{eq:GF0} and~\eqref{eq:GFl} into Eq.~\eqref{eq:K1def}.
Rearranging the terms in order to explicitly cancel out the on-site interaction terms, we get
\begin{multline}
\cm_{1} = \beta\int \frac{dpdp'}{(2\pi)^{2}} \left(\bar{V}_{p'-p} - \bar{V}_{0} \right)
\times\\\times
 n_{p'} \sum_{kl} (1-n_{p})n_{p}
e^{ip(l-k)} (D^{-1})_{kl}
+\\
+\int \frac{dpdp'dq}{(2\pi)^{3}} \left(\bar{V}_{p'-p-q} - \bar{V}_{q}\right)
W_{pp'q}
\sum_{mnkl}\\\
e^{-ip'm} (D^{-1})_{mn}e^{i(p'-q)n}
e^{-ipk} (D^{-1})_{kl} e^{i(p+q)l}
\end{multline}
where $\bar{V}_{q} = \sum_{n=-\infty}^{\infty} V_{l}e^{iql}$ is simply a Fourier coefficient and
\begin{equation}
W_{pp'q} =
\frac{n_{p}(1-n_{p+q})n_{p'}(1-n_{p'-q})}%
{-\xi_{p'} + \xi_{p'-q} - \xi_{p} + \xi_{p+q}}.
\end{equation}
For the case of the nearest neighbor interaction the Fourier series for the interaction potential takes simple form $\bar{V}_{q} = V_{0} + 2V_{1}\cos q$, where $V_{0}$ is an irrelevant on-site interaction that cancels out in the given above formula for $\cm_{1}$.

\section{Approximating sine-Gordon model via discrete harmonic oscillator chain}
\label{apx:wigner}
\setcounter{figure}{0}

\begin{figure}
\centering
\includegraphics[scale=0.7]{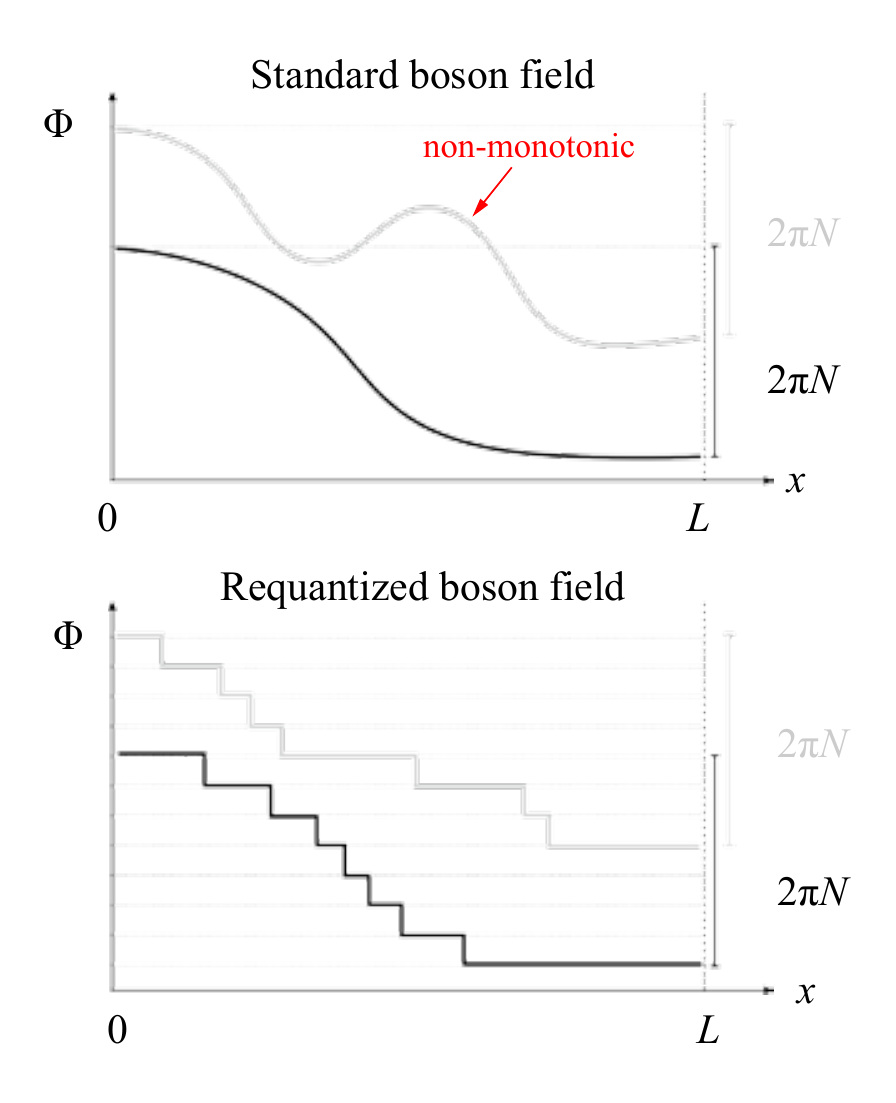}
\caption{The principle of a requantization of the boson field $\Phi$.
(a) Possible configurations of $\Phi$ as a function of x for standard Luttinger liquid theory.
The black and gray curve represent possible quantum superpositions of two different realizations of $\Phi$.
Only the total difference $\Phi(0)-\Phi(L)$ is an integer multiple of $2\pi$, but for arbitrary positions, $\Phi(a)-\Phi(b)$ the field may assume arbitrary values.
Thus, any local charge operator is not quantized in general.
(b) Configuration of $\Phi$ including a mass term in the nonrelativistic limit, $c\to \infty$, again for two different quantum realizations (black and gray).
Here, the mass term leads to a step-like behavior, such that $\Phi(a)-\Phi(b)$ is an integer multiple of $2\pi$ independent of $a$ and $b$.
Thus, the local charge is always guaranteed to be quantized.
}
\label{fig:q2bfield}
\end{figure}

We here reiterate a justification for the Wigner crystal model for interacting electrons, starting from the sine-Gordon model.
Subsequently, we use the former to compute the charge noise, given in the main text in Eq.~\eqref{eq:wigner}.
As shown by Coleman and Mandelstam~\cite{Coleman_1975,Mandelstam_1975}, the massive Thirring model (interacting relativistic fermions in $1+1$ dimensions) can be exactly mapped onto the sine-Gordon model.
In simple words, we can take the Luttinger liquid Hamiltonian and add a mass term,
\begin{equation}
\label{eq_sine_Gordon}
\!\!\!\!\!H\!=\!\!\int \!\! dx\left[\frac{cK}{2}\Pi^{2}\!+\!\frac{c}{2K}\left(\partial_{x}\Phi\right)^{2}\!-\!\frac{\alpha}{4\pi K}\cos\left(2\sqrt{\pi}\Phi\right)\right]\!,\!\!\!
\end{equation}
where $K$ parametrizes the interactions, and a nonzero $\alpha$ gives rise to a mass gap.
Let us treat the soliton positions of the sine-Gordon model given in Eq.~\eqref{eq_sine_Gordon} semiclassicaly, as a chain
of harmonic oscillators.
Assuming a non-relativistic limit, i.e., when the cosine terms dominate, the smooth field $\Phi$ will be looking rather like a a staircase, see Fig.~\ref{fig:q2bfield}.
The dynamics of the system will be rather described by the positions of the kinks $x_{j}$, the simplest model for which will be an Eq.~\eqref{eq:oscillatorschain}.
To quantize the oscillators chain we deploy the ansatz
\begin{align}
\widetilde{x}_{j} &= x_{j}-j\frac{L}{N} \nonumber \\
& =X+\frac{1}{\sqrt{2N}}\sum_{q>0}\frac{1}{\sqrt{\alpha_{q}}}\left[e^{iqj}a_{q}+e^{-iqj}a_{q}^{\dagger}\right] \\
p_{j} & =\frac{1}{N}P-i\frac{1}{\sqrt{2N}}\sum_{q>0}\sqrt{\alpha_{q}}\left[e^{iqj}a_{q}-e^{-iqj}a_{q}^{\dagger}\right]
\end{align}
where $\left[a_{q},a_{q'}^{\dagger}\right]=\delta_{qq'}$, momentum $q=\frac{2\pi}{N}n\quad n\in\mathbb{Z}$, while $X=\frac{1}{N}\sum_{j=1}^{N}x_{j}$ and  $P=\sum_{j=1}^{N}p_{j}$.
The normalization factor has to be chosen as $\alpha_{q}=2 m\omega\left|\sin\left(\frac{q}{2}\right)\right|$.
Adding the chemical potential we get the diagonalized Hamiltonian
\begin{align}
H=\frac{1}{2m}\frac{1}{N}P^{2}+\omega\sum_{q>0}2\left|\sin\left(\frac{q}{2}\right)\right|a_{q}^{\dagger}a_{q}+
\nonumber\\
+\omega\sum_{q>0}\left|\sin\left(\frac{q}{2}\right)\right|+\frac{m\omega^{2}}{2}\frac{L^{2}}{N}-\mu N.
\end{align}
Comparing this with the bosonized Luttinger liquid Hamiltonian in the diagonalized
form
\begin{equation}
H_{LL}=u\sum_{q\neq0}\left|q\right|a_{q}^{\dagger}a_{q}+\frac{\pi u}{2L}\left(\frac{\left(N-N_{F}\right)^{2}}{K}+KJ^{2}\right)
\end{equation}
we can obtain all necessary values 
\begin{equation}
\omega = v_{F}\frac{N}{L},\quad k_{F}=\pi\frac{N}{L},\quad m=\frac{\pi}{v_{F}}\frac{N}{L}.
\end{equation}
Putting the expressions for $x_{i}$ into Eq.~\eqref{eq:Q1quant} we expand over $\delta\widetilde{x}_{j}^{r}$ as follows
\begin{align}
Q \!&=\! \sum_{j} \! \frac{1}{\pi} \! \left[\theta\left(\! \frac{L}{N}j \!+\! X \!+\! \delta\widetilde{x}_{j} \!\right) \!-\!\theta\left(\! \frac{L}{N}j \!+\! X \!+\! \delta\widetilde{x}_{j} \!-\! l \!\right)\right]
\nonumber\\&=
\sum_{rj}\frac{1}{\pi}\frac{\delta\widetilde{x}_{j}^{r}}{r!}\partial_{X}^{r}\left[\theta\left(\! X \!+\! \frac{L}{N}j \!\right) \!-\! \theta\left(\! X \!+\! \frac{L}{N}j \!-\! l \!\right)\right]\!.
\end{align}
The derivatives of the $\theta$-functions can be obtained using
$$
\frac{1}{\pi}\left[\theta(a)\!-\!\theta(b)\right] \!=\! \frac{1}{2\pi}\!\int\! dk \!\int_{a}^{b}\! dX e^{ikX} \!=\! \frac{1}{2\pi} \!\int\! dk \frac{e^{ikb}\!-\!e^{ika}}{ik}.
$$
To calculate first $\left\langle Q\right\rangle$ and second $\left\langle Q^{2}\right\rangle$ moments we need $\left\langle \delta\widetilde{x}_{j}^{r}\right\rangle $ and $\left\langle \delta\widetilde{x}_{j}^{r}\delta\widetilde{x}_{j'}^{r'}\right\rangle $ which can be computed via generating function equal to (for large $N\gg1$)
\begin{gather}
\left\langle e^{i\xi\delta\widetilde{x}_{j}}\right\rangle  \approx e^{-\xi^{2}a},
\\
\left\langle e^{i\left(\xi\delta\widetilde{x}_{j}+\zeta\delta\widetilde{x}_{j'}\right)}\right\rangle =e^{-K\left[\left(\xi+\zeta\right)^{2}a-2\xi\zeta b\left(j-j'\right)\right]}
\end{gather}
where
\begin{align}
a  &\!=\!-\frac{1}{8\pi^{\frac{4}{3}}}\frac{L^{2}}{N^{2}}\ln\left(\frac{\pi}{2N}\right)
\\
b\left(j\!-\!j'\right) &\!=\!
\begin{cases}
\frac{1}{2\pi^{2}}\frac{L^{2}}{N^{2}}\left[1\!+\!\frac{1}{2}\ln\left|j\!-\!j'\right|\right]&\text{for }j\neq j'\\
0&\text{for }j=j'
\end{cases}
\end{align}
The calculation of the charge gives us $\left\langle Q\right\rangle =l\frac{N}{L}$.
The noise is given through an expression
\begin{align}
&\left\langle Q^{2}\right\rangle = \frac{N}{2L}\sum_{\delta j}\int_{0}^{l}dX
\times \nonumber\\&\times
\left[\text{erf}\left(\frac{l-\frac{L}{N}\delta j-X}{2\sqrt{2Kb\left(\delta j\right)}}\right)-\text{erf}\left(\frac{-\frac{L}{N}\delta j-X}{2\sqrt{2Kb\left(\delta j\right)}}\right)\right].
\end{align}

\section{Convergence of the perturbation series}
\label{apx:convergence}
\setcounter{figure}{0}

The convergence of the perturbation series can be violated in two ways.
The first way is the divergency due to the momenta integration.
The parameter $L$ serves as a natural cutoff for this integration, however, since we are interested in $L\gg1$, let us consider this integration for arbitrary order in interaction expansion.
The perturbative term of the order $M$ to the cumulant of order $P>1$ can be estimated formally as follows.
For $M$ interaction vertexes we have $2M$ Green's functions, $P$ of which are correction of the dressed function $\tilde{G}$,
so in result we have $2M+P$ bare Green's functions $G^{(0)}$ estimated as $(i\omega_{n}-\xi)$ and $P$ additional momenta $q$ together with prefactors $q^{-1}$ and typical cutoffs at $k_{F}\lesssim |q|\lesssim L^{-1}$.
Except for that we have initial $2M$ integrations over momenta, $2M+P$ summations over the Matsubara frequency $\omega_{n}$, and $M$ conservation laws for both momenta and frequency.
Putting all of this together we get
\begin{equation}
V^{M}\frac{\zeta^{P}}{\xi^{M}}
\left(\int dp\right)^{M}
\left(\int_{k_{F}>|q|>L^{-1}}\frac{dq}{q}\right)^{P}
\end{equation}
As we see, potentially the maximal divergence we may obtain in any perturbation order is $\ln k_{F}L$.

\begin{figure}
\centering
\includegraphics[width=.9\columnwidth]{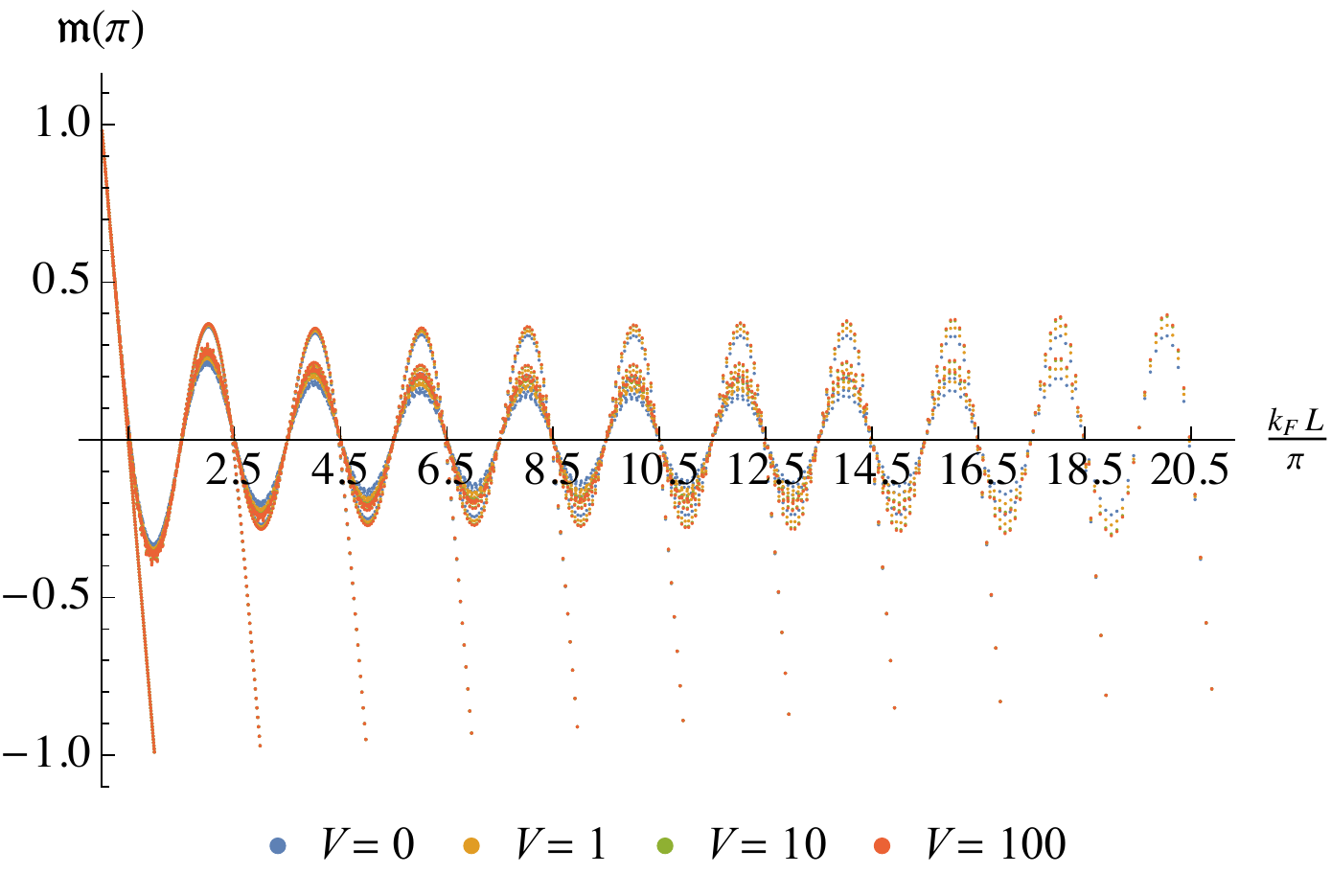}
\caption{DMRG computations of the parity expectation values.
The color designates the interaction strength.
The data points are collected for $k_{F}$ from $\pi/20$ to $\pi/10$, the interval length varies in $L\in[1..199]$ at total number of sites $200$.
}
\label{fig:plotMpidmrg}
\end{figure}

The other source of the probable non-analytic behavior is the Green's function itself.
As we mentioned in the end of Section~\ref{sec:diags}, the denominator in the Green's function definition is equal to $e^{\cm_{0}}$ and goes to zero at $\lambda=\pi$ and $k_{F}=k_{C,n}$.
Thus, the diagram of the $M^\text{th}$ order can be estimated as $V^{M}\times(k_{F}-k_{C,n})^{2M}$, and, obviously, the diagrammatic expansion in small $V$ fails if $k_{F}$ approaches $k_{C,n}$.
For this reason we performed a DMRG computation of the parity expectation value (i.e., generating function at $\lambda=\pi$) for the chain of length $N_\text{sites}=200$ with boundary periodic conditions for different $k_{F}$, $L=1..N_\text{sites}-1$, and various interaction strengths.
The result is presented in Fig.~\ref{fig:plotMpidmrg}.
We observe that the vicinity of the $n+1/2$ points, where our perturbative approach fails, is the least affected as by interaction strength (in \emph{very} wide range), so by the particulars values of $k_{F}$ and $L$ for the given $k_{F}L = (n+1/2)\pi$ value.
Thus, we can conclude that the interaction correction $\cm(\pi)-\cm_{0}(\pi)$ remains finite in the limits of $k_{F}\to k_{C,n}$.


\bibliographystyle{apsrev4-2}
\bibliography{counstat}

\end{document}